# Vertex Reconstructing Neural Network at the ZEUS Central Tracking Detector

Gideon Dror[1+], Erez Etzion[2*]

1. Department of Computer Science, The Academic College of Tel-Aviv-Yaffo, Tel-Aviv 64044, Israel.
2. School of Physics and Astronomy, Raymond and Beverly Sackler Faculty of Exact Sciences, Tel-Aviv University, Tel-Aviv 69978, Israel.

**Abstract.** An unconventional solution for finding the location of event creation is presented. It is based on two feed-forward neural networks with fixed architecture, whose parameters are chosen so as to reach a high accuracy. The interaction point location is a parameter that can be used to select events of interest from the very high rate of events created at the current experiments in High Energy Physics. The system suggested here is tested on simulated data sets of the ZEUS Central Tracking Detector, and is shown to perform better than conventional algorithms.

## INTRODUCTION

The $z$ coordinate of the interaction vertex position in collider experiments can be used for a prompt decision whether to record the event or reject it. Here we explore an unconventional neural network (NN) approach to vertex finding tested with simulation of data collected by the ZEUS[1] detector at HERA. A collision of electron and proton occurs at HERA every 96 nsec. A very small rate of signal is selected from the background by a three-level triggering system that performs filtering through software or hardware. The challenge is to find an efficient way to extract the location of the interaction vertex along the $z$-axis, determined by the direction of the colliding protons. In the following we use simulated Central Tracking Data (CTD)[2] events and compare our results with those obtained by the second level trigger. Such a NN calculation may be very fast when implemented on dedicated hardware.

The basic idea is to use a network with fixed architecture that is inspired by the way our brain processes visual information. It is similar to the orientation selective NN employed by [3] which was used to select linear tracks. The present network is based on our identification method. Previous successful attempt[4] required a network size for hardware realization. Here we show that a much smaller network can be used without any precision loss.

## ZEUS CTD

The ZEUS CTD, placed in a magnetic field of 1.43T, surrounds the interaction point and is designed to measure direction, charge and momentum of charged particles. It has a cylindrical shape around the $z$-axis. 4608 signal wires are organized in 9 super-layers; each consists of 8 wire layers. Wires in odd super-layers run parallel to the $z$-axis while $5^0$ tilt wires in even ones. The three inner layers provide also a $z$ position by comparing the pulse time of arrival at both ends of the chamber.

∴ Presentes at ACAT 2000, FermiLab, Chicago, October 2000
+ Gideon@server.mta.ac.il
* Erez@lep.tau.ac.il

## THE INPUT DATA

The input consist of the (*x,y*) positions of hits in the axial super-layers (1,3,5,7,9) and the *z* timing for hits in the super-layers 1,3,5.

Fig. 1 shows an example of an event projected onto *z*=0 plane. Several slightly curved tracks (arcs) starting at the origin can be seen, each of which is made of 30-40 data points.

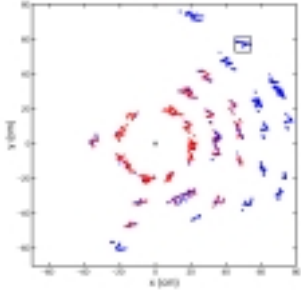

**FIGURE 1.** A typical event projected into the *z*=0 plane. The dots represent hits in the CTD. Red dots contain *z* information.

Since the tracks are expected to form a helix shape with small curvature, one expects a linear dependence of the z coordinates of the hits on their radial position. It was demonstrated that trying to fit the hits to a straight-line resulted in a considerable scatter of the data.

## THE NETWORK

The network is based on step-wise changes in the representation of the data, moving from the input points, to local line segments to local arcs and finally to global arcs. Two parallel calculations deal separately with the two problems: *xy* and *z*-coordinates. The first NN, which handles the *xy* information, is responsible for constructing the arcs that correctly identify some of the particle tracks in the event. The second process uses the information to evaluate the *z* location of the point where all tracks meet.

The arc identification NN processes information in the fashion akin to the way visual information is processed by the primary visual system [5]. The input layer is made of a large number of neurons. The total area to cover in *xy* is 5000 cm$^2$ and we initially covered it with 100000 input neurons[4].

Neurons of the first layer are line segment detectors labeled by (*XY*α) – the coordinates of the segment center and its orientation. The activation of the first layer is given by $V_{XY\alpha} = g(\sum_{XY} J_{XY\alpha,xy} V_{xy} - \vartheta_1)$ where *J* is 1 when the distance between *xy* and *XY* < 0.5, -1 when 0.5<distance<2, and zero elsewhere. Fig. 2 (Left) represents the output of this layer.

Neurons of the second layer transform the representation of local line segments to local arc segments. They are defined by (κ,θ,i) their curvature κ=1/R, θ, the slope at the origin and i which relates each arc segment to the super-layer it belongs to. The mapping between the second and third layers is done in the following way: For a given local arc segment, we take the arc segment which is closest to it.

The neurons belonging to the last layer are global arc detectors, and detect projected tracks on to the z=0 plane. These neurons are denoted by (κ,θ) and are connected to the second layer in a simple fashion, $V_{\kappa\vartheta} = g(\sum_{\kappa`\vartheta`i} \delta_{\kappa,\kappa`} \delta_{\vartheta,\vartheta`} V_{\kappa`,\vartheta`,i} - \vartheta_3)$. Fig. 2 (Right) represents the activity of the third layer neurons. Each active neuron is equivalent to an arc in the figure.

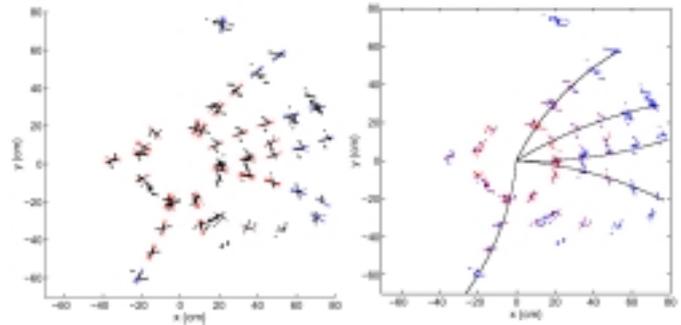

**FIGURE 2.** Representation of the activity of (Left) first layer neurons *XY*α for Fig. 1 input. At some of the points several line segments with different directions occur due to the low threshold (θ$_l$). On the right last layer neurons plotting the appropriate arcs in κθ plane.

The input to vertex *z* location finding is the mean value of points within the receptive field of the first NN input neurons.

The first layer neurons compute the mean value of the *z* coordinate of the first layer neurons in their receptive field, averaging over all the neurons within the section. This is similarly propagated to the second layer neurons.

The third layer neurons evaluate the *z* value of the arcs origin by simple extrapolation. The final *z* estimate of the vertex is calculated be averaging the output of all active third layer neurons.

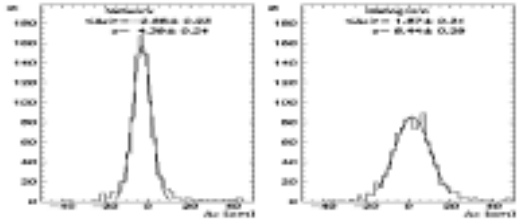

**FIGURE 3.** A distribution of the error in the estimation of the z values, the NN estimate on the left and the conventional histogram method on the right.

Fig. 3 demonstrates how the NN $z$ estimate is more precise than the conventional histogram method [6].

## RESULTS

The results of the network were tested with 992 simulated HERA events, comparing the results to the actual location of the vertex $z$ as well as to the histogram method. There is a large overlap between events with no estimation. However, the resolution of the NN method ($\sigma \sim 4.5$ cm) is by almost a factor of two better than by the conventional method ($\sigma \sim 8.4$ cm). ZEUS full off-line reconstruction of leads to a resolution of 0.1 cm. Taking the number of output neurons to represent the number of track candidates passing through the vertex, we compared it to the histogram method and to the number of tracks in the full off-line reconstruction. Both methods were comparable and both tend to slightly underestimate the number of tracks from the vertex.

More than 300 NN differing in the number of neurons and their configurations, were examined. Fig 4. shows that we could decrease the number of input neurons as well as the resolution in the second layer without significantly affecting the performance of the network. Lateral connection between 1$^{st}$ layer neurons enabled us to reduce the threshold and by that to reduce the network size

## DISCUSSION

We have described a feed-forward NN that performs a task of pattern identification by threshold and data subsets selection. The network uses a fixed architecture; this allows a hardware implementation, which may be crucial for fast triggering purposes. The pattern type we were after motivated the fixed architecture and the present synaptic connections. The obtained results are better than conventional methods, which enable the opportunity of new NN implementation in triggering devices of HEP experiments.

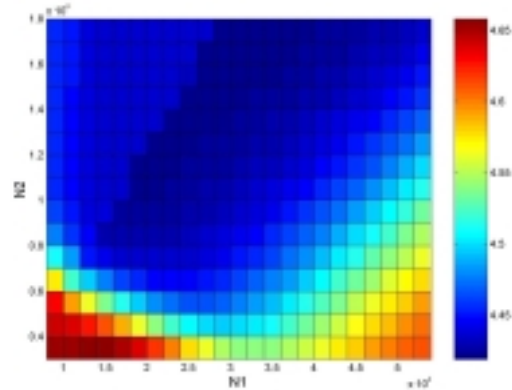

**FIGURE 4.** The network output width as a function of the number of neurons in the input layer (N1 axis) and second layer (N2 axis). (Blue- smaller $\sigma$, Red – Larger $\sigma$)

## ACKNOWLEDGMENTS

The study has been partly pursued in the framework of the ZEUS Collaboration and we would like to acknowledge the help of our colleagues Halina Abramowicz from ZEUS in providing the necessary data sample and for stimulating discussions. We wish to thank David Horn from Tel-Aviv University for the fruitful collaboration. This research was partly supported by the Israeli National Science Foundation.

## REFERENCES

1. ZESU Collab., The ZEUS Detector, Status Report 1993, DESY 1993; M. Derrick et al.,. *Phys. Lett.* **B293**, 465 (1992).

2. B. Foster et al., *Nucl. Inst. Meth.* **A338**, 254 (1994).

3. H. Abramowicz, D. Horn, U. Naftaly and C. Sahar-Pikielny, *Nucl. Inst. Meth.* **A378**, 305 (1996); *Advanced in Neural Information Processing Systems 9*, edited by M.C. Mozer, M.J Jordan and T. Petsche, MIT Press, 1997, pp. 925.

4. G. Dror et al., *AIHENP99 proceedings*, edited by G. Athanasiu, D. Perret-Gallix, Elsevier North-Holland Editors, (1999).

5. D.H. Hubel ,T.N Wiesel, *J. Physiol.*. **195**, 215 (1968).

6. Quadt, Master of Science Thesis, University of Oxford (1997).